\documentclass[conference]{IEEEtran}
\usepackage{booktabs}
\usepackage{amsfonts}
\usepackage{amssymb}
\usepackage{amsmath}
 
\usepackage{graphicx}
\usepackage{subfigure}
\usepackage{cite}
\usepackage{algorithm}
\usepackage{textcomp}
\usepackage{algpseudocode}
\usepackage{color,soul}
\usepackage{amsfonts}
\usepackage{epstopdf}
\usepackage{multirow} 

\begin{document}
\title{Active User Detection of Uplink Grant-Free SCMA in Frequency Selective Channel}
\author{Feilong Wang, Yuyan Zhang, Hui Zhao, Hanyuan Huang, Jing Li\\
Intelligent Computing and Communication Lab, BUPT, Beijing, China\\
Key Laboratory of Universal Wireless Communication, Ministry of Education\\
Email: flwang@bupt.edu.cn
\vspace*{-3pt}
}

\maketitle
\begin{abstract}
Massive machine type communication (mMTC) is one of the three fifth generation mobile networking (5G) key usage scenarios, which is characterized by a very large number of connected devices typically transmitting a relatively low volume of non-delay sensitive data. To support the mMTC communication, an uplink (UL) grant-free sparse code multiple access (SCMA) system has been proposed. In this system, the knowledge of user equipments' (UEs') status should be obtained before decoding the data by a message passing algorithm (MPA). An existing solution is to use the compressive sensing (CS) theory to detect active UEs under the assumed condition of flat fading channel. But the assumed condition is not suitable for the frequency selective channel and will decrease the accuracy of active UEs detection. This paper proposes a new simple module named refined active UE detector (RAUD), which is based on frequency selective channel gain analyzing. By making full use of the channel gain and analyzing the difference between characteristic values of the two status of UEs, RAUD module can enhance the active UEs detection accuracy. Meanwhile, the addition of the proposed module has a negligible effect on the complexity of UL grant-free SCMA receiver.


\end{abstract}
\begin{IEEEkeywords}
Sparse code multiple access; Uplink grant-free; Active user equipments detection

\end{IEEEkeywords}
\section{Introduction}
\IEEEPARstart{F}{ifth} generation mobile networking (5G) is expected to support a scenario with low latency and massive connectivity, such as the one of the three key usage scenarios Massive machine type communication (mMTC). While the current long term evolution (LTE) system is not efficient enough, especially in the uplink (UL) multi-user access scenario. 

Sparse code multiple access (SCMA) \cite{b1} as a new multiple access technology for massive connectivity, can enhance the maximum number of accessible user equipments (UEs) in wireless channels.
The incoming data streams are mapped to codewords of different multi-dimensional codebooks and these sparse codebooks can share the same time-frequency resources of orthogonal frequency division multiple access (OFDMA). 
Due to the sparse characteristic of codewords, receiver can use the message passing algorithm (MPA), which can achieve near-optimal detection with low complexity.

To reduce the transmission latency, UL grant-free transformation is proposed, which also can reduce the overhead associated with control signals for scheduling. In UL grant-free multiple-access scenario, UEs are allowed to transmit data in pre-scheduled resources at any time. The pre-scheduled resources is called contention transmission unit (CTU), which  includes time, frequency, codebooks for active UEs \cite{b2}. So this requires the receiver to be able to detect active UEs without the knowledge of active codebooks and pilots, to estimate their fading channels, and to decode their data, .

In \cite{b3}, an active UE detector (AUD) module has been proposed to detect active UEs and reduce the number of potential active UEs. By utilizing the orthogonality of pilots of different UEs and using the received pilot signal, AUD module can identify the status of UEs. Then, an joint message passing algorithm (JMPA) was proposed to eliminate the false detected inactive UEs caused by AUD module and decode the data of active UEs. Despite the AUD module has a not very high active UEs detection accuracy, but it can cut down the complexity of JMPA.

The pilots of CTU are used as a known condition at the receiver, and they are always calculated in form of a matrix. Because the pilot matrix satisfies the restricted isometry property (RIP) with a high probability and the signal model has sparse characteristic for AUD module, active UEs detection can be solved well through compressive sensing (CS) theory. The classical CS theories are compressive sampling matching pursuit (CoSaMP) \cite{b4} and iterative support detection (ISD) \cite{b5}. 
A modified version of the original ISD algorithm called the structured ISD (SISD) was proposed in \cite{b6}, which can jointly detect the active UEs and the received data in several continuous time slots. 
In \cite{b7}, a novel sparsity-inspired sphere decoding (SI-SD) algorithm was proposed to integrate AUD into JMPA modules, with a lower computational complexity avoiding the redundant pilot overhead. However, the above two algorithms in \cite{b6} and \cite{b7} are assume that the channel gain of all potential UEs is already know without the discussion of the way to get them. Actually the receiver can not estimate channel gain of the inactive UE who transmit noting to receiver. 
In \cite{b8}, \cite{b9}, an algorithm based on the framework of sparse bayesian learning (SBL) is proposed to reduce the requirement of pilots overhead and improve active UEs detection performance for AUD module. 
Despite the practical scenario is under frequency selective channel, the SBL algorithms of AUD module is still realized based on the assumption that active UEs go through a flat fading channel. So the active UEs detection performance of the AUD algorithms will become inaccurate.

In this paper, a new refined AUD (RAUD) module is proposed to enhance the active UEs detection accuracy of the receiver in an UL grant-free SCMA system. By making full use of the channel gain and analyzing the difference between characteristic values of the two status of UEs, the RAUD module makes the receiver have better active UEs detection accuracy and adaptability in frequency selective channel. In Section II, we introduce an original scheme of the grant-free SCMA access model, then analyze the principle, advantages and disadvantages of AUD, channel estimator (CE) and JMPA module. Section III is devoted to describe the RAUD module and the two-step AUD receiver contains RAUD module. Numerical results are provided in Section IV to evaluate the performance of the proposed two-step AUD receiver in UL grant-free SCMA scenario, and the concluding remarks are given in Section V.

\section{SYSTEM MODEL}

\subsection{UL Grant-free SCMA Transmitter}

\begin{itemize}
	\item SCMA encoder
\end{itemize}

An SCMA decoder can be seen as a mapper from $\log _2 M$ bits to a codeword in the $T$-dimensional complex codebook with size $M$. The $T$ dimensional codeword of the codebook is a sparse vector with $N$ non-zero elements, i.e., $N<T$. Then, different UEs' codewords will share the same time-frequency resources to be transmitted.

\begin{itemize}
	\item Codebooks and pilots selection
\end{itemize}

In the UL grant-free SCMA scenario, a basic resource called CTU is known for transmitter and receiver, which is defined as a combination of SCMA codebooks for encoding data and detecting active UEs, pilots for identifying UEs and estimating channel, and time-frequency resource. As shown in Fig. 1 (a), over a time-frequency resource, there are $J$ groups, each of them includes one codebook $\bf{C}$ and $L$ pilots $\bf{S}$. The codebook in each group corresponds to the same Zadoff-Chu (ZC) sequence, but with different cyclic-shifts on the sequence to generate different pilots\cite{b10}\cite{b11}. The location of the pilot in OFDM time–frequency grid can reference LTE UL Demodulated Reference Signal (DMRS) \cite{b12}\cite{b13}, as show in Fig. 1 (b). Hence, a total $K=J \times L$ unique pilots $\{ {\bf{s}}_1, \cdots , {\bf{s}}_K \}$, called pilot pool, is pre-defined.

\begin{figure}[!t]
\centering
\includegraphics[width=0.49\textwidth]{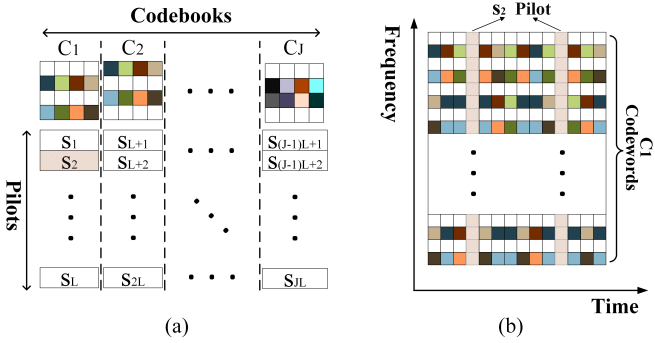}
\vspace*{-8pt}
\caption{(a) Definition of CTU; (b) an example of signaling for a grant-free random access user who use ${\bf{s}}_2$ pilot and  corresponding ${\bf{C}}_1$ codebook.}
\label{1}
\vspace*{-5pt}
\end{figure}

In a cell with a large number of UEs, if small subsets of these UEs have data streams to be transmitted, each of them has to randomly select a pilot from the pilot pool and a corresponding codebook from the CTU. Then, the data streams can be mapped to codewords and transmitted with the pilot, represented in Fig. 1 (b). These UEs who transmit data are called active UEs, while other UEs are called inactive UEs. Therefore, the receiver of the UL grant-free SCMA system has to decode data of active UEs without the knowledge of UEs' status, codebooks and pilots they selected.

\subsection{UL Grant-free SCMA Receiver}

In \cite{b3}, the UL grant-free SCMA receiver structure can be shown as Fig. 2. There are three modules in the receiver structure, i.e., AUD, CE and JMPA. The input are the received pilot signal and the received data signal. The output is the decoded data of the active UEs.

\begin{figure}[!t]
	\centering
	\includegraphics[width=0.47\textwidth]{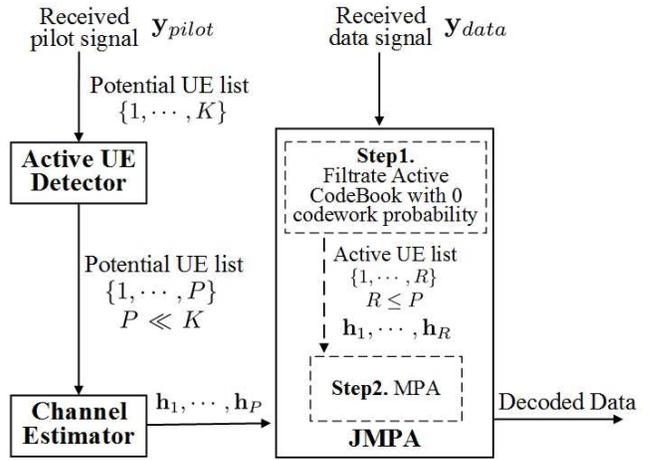}
	\vspace*{-8pt}
	\caption{UL grant-free SCMA receiver structure including active UE detector, channel estimator and JMPA.}
	\label{2}
	\vspace*{-5pt}
\end{figure}

\begin{itemize}
	\item Active UE Detector
\end{itemize}

The AUD module is able to identify active UEs/pilots and reduce the size of the potential UE list form $K$ to $P$, $P \ll K$. The received pilot signal model for AUD is represented in Eq.(1), where ${\bf{y}}_{pilot} = [ y_{1}, y_{y2}, \cdots , y_{Q} ] ^{\rm T} $ is the received pilot signal, $Q$ is the length of the sequence. ${\bf{S}}$ is the pilot matrix containing $K$ potential pilot sequences. ${\bf{s}}_k = [ s_{k1}, s_{k2}, \cdots , s_{kQ} ] ^{\rm T}$ is the $k$-th pilot sequence. Each element of the sparse signal ${\bf{h}}$ corresponds to one pilot. $ {\bf{n}} = [ n_1, n_2, \cdots , n_Q ] ^{\rm T}$ is the noise vector following the distribution $ CN\left( {0,{{\sigma}^2}{{\bf{I}}_N} }\right)$.

\begin{equation}\label{1}
\begin{aligned}
{\bf{y}}_{pilot} &= {\bf{S}}{\bf{h}} + {\bf{n}}  \\
&=\left[\begin{array}{*{20}c}
{s_{11} } & {s_{21} } &  \cdots  & {s_{K1} }  \\
s_{12} & {s_{22} } &  \cdots  & {s_{K2} }  \\
\vdots & \vdots &  \ddots  & \vdots  \\
s_{1Q} & {s_{2Q} } &  \cdots  & {s_{KQ} }  \\
\end{array} \right]\cdot 
\left[{\begin{array}{*{20}c}
	{h_1}  \\
	{h_2}  \\
	\vdots  \\
	{h_K}  \\
	\end{array}} \right]
+ {\bf{n}}
\end{aligned}
\end{equation}
\vspace{0.05mm}

The active UEs/pilots detection problem can be seen as compressed sensing for the sparse characteristic of ${\bf{h}}$ in Eq.(1). After compressed sensing algorithm, such as focal underdetermined system solver (FOCUSS)\cite{b14}, the status of $k$-th pilot/codebook can be identified by the size of $\lvert h_k \rvert^2$. We call $\lvert h_k \rvert^2$ characteristic value of $k$-th UE. If $\lvert h_k \rvert^2 \geq \lambda_{AUD}$, the $k$-th UE/pilot is declared to be active. The value of $\lambda_{AUD}$ is usually set to 0.01.
\vspace{1mm}


\begin{itemize}
	\item Channel Estimator
\end{itemize}

The CE module can estimate channel gain from the received pilot signal ${\bf{y}}_{pilot}$ and the $P$ potential active pilot sequences. The received pilot signal model for CE module is represented in Eq.(2). 

\begin{equation}\label{2}
{\bf{y}}_{pilot} = \sum\limits_{p = 1}^P { diag({\bf{h}}_p){\bf{s}}_p} + {\bf{n}} 
\end{equation}
\vspace{0.05mm}

Although Eq.(2) is the same as Eq.(1) in regard to the received pilot signal ${\bf{y}}_{pilot}$, there are still some difference. Firstly, there is a total of $P$ potential active pilot sequences left after AUD module. Secondly, each pilot sequence ${\bf{s}}_k = [ s_{p1}, s_{p1} \cdots , s_{pQ} ] ^{\rm T}$ corresponds to one channel gain vector ${\bf{h}}_p = [ h_{p1}, h_{p2}, \cdots , h_{pQ} ]$ in pilot location of $p$-th UE. The mature channel estimation algorithm, such as  Minimum Mean Square Error (MMSE), has a better performance.
%


For a frequency selective channel, the symbols in a pilot sequence over different subcarriers will correspond to different fading values. Therefore, comparing $\bf{h}$ in Eq.(1) with $h$ in Eq.(2), the received pilot signal model of CE module is more accurate than the model of AUD module. 
But meanwhile, there are two advantages that AUD module use Eq. (1) as the calculation model. Firstly, the model can be solved well through mature CS theory. Secondly, it makes algorithm complexity lower than the model in Eq.(2) does, because the number of pilot sequences in pilot pool $K$ is a big orders of magnitude.

\begin{itemize}
	\item JMPA
\end{itemize}

On account of the small-scale deep fading channel and signal noise, the inaccuracy of AUD module is difficult to avoid. Then, in \cite{b3}, the function of JMPA algorithm is proposed to identify the false detected inactive UEs further and decode the data of the real active UEs. The received data signal vector ${\bf{y}}_{data} = [ y_{1}, y_{y2}, \cdots , y_{Q} ] ^{\rm T} $ for JMPA is obtained from Eq.(3). 

\begin{equation}\label{2}
{\bf{y}}_{data} = \sum\limits_{p = 1}^P { diag({\bf{g}}_p){\bf{x}}_p} + {\bf{n}} 
\end{equation}
\vspace{0.05mm}

\noindent
where ${\bf{g}}_p$ is the channel gain in data location and ${\bf{x}}_p$ is the codeword symbols for $p$-th active UE. $ {\bf{n}}$ is the noise vector following the distribution $ CN\left( {0,{{\sigma}^2}{{\bf{I}}_N} }\right)$.

JMPA algorithm specifically contains two steps:

\textbf{Step1}: Identify inactive UEs and remove them from the potential UE list. If an inactive UE has no codewords mapped from data stream to transmit, it is equivalent to transmit zero codewords, whose values are zero. In fact, JMPA can be considered as MPA which use the new codebook contains zero codeword. Therefore, JMPA can identify inactive UEs from the probability of zero codeword and nonzero codeword. We may expect that the probability of zero codeword for inactive UEs is much higher than that for active UEs. This gap enable JMPA to identify the status of UEs.

\textbf{Step2}: Use the MPA algorithm to decode the data of the real active UEs.

In the practical simulation for JMPA module, there are some factors that make the active UEs detection performance not as ideal as the above expectation. 
Firstly, the transmission power of inactive UEs, who transmit zero codewords, is much lower than that of active UEs. Secondly, the channel gain of inactive UEs, whose pilot sequence is not included in the received pilot signal in Eq.(2), is lower than that of active UEs. Meanwhile the JMPA module consider that the noise power of inactive UEs is same as that of active UEs. 
So, the equivalent signal noise ratio (SNR) of inactive UEs is lower than that of active UEs. For JMPA, the probability of zero codeword of inactive UEs will be close to or even lower than that of active UEs. That makes it hard to identify the status of UEs.

\section{TWO-STEP AUD RECEIVER FOR UL GRANT-FREE SCMA SYSTEM}

In Section II, the principle, advantages and disadvantages of AUD module have been analyzed. Then we intend to use the implicit information in received pilot signal, which is neglected by AUD module. And in this way, the inactive UEs misinterpreted by AUD module can be selected out.

Observing Eq.(1), the received pilot signal model of AUD module, and Eq.(2), the received signal model of CE module. The received pilot signal  ${\bf{y}}_{pilot}$ doesn't contain the pilot information of inactive UEs. And there is a strong correlation between pilots $\bf{s}$ of different UEs. Therefore, in AUD module, the estimated coefficient $h$ of $\bf{s}$ for inactive UE tends to zero. Also, in CE module, the estimated coefficient $\bf{h}$ of $\bf{s}$ for inactive UE tends to zero vector. In other words, from the receiver's point of view, an inactive UE can be considered as a UE who goes through an infinitely deep fading channel. It is analyzed in Section II that $\bf{h}$ contains more subcarrier information than $h$ and can accurately reflect the channel information. Therefore, $\bf{h}$ can be used to select out the inactive UEs that AUD module misinterpreted. Then, we propose a RAUD module that takes the channel gain vector $\bf{h}$ as input. By using the difference between active UEs and inactive UEs about $\bf{h}$, the RAUD module can output a refined active UE list.

In order to distinguish the $\bf{h}$ of the two status of UEs, we define a metric called $\bf{F}$ characteristic value. The value of $\bf{F}$ is obtained from 1-norm $||\bf{h}||_1$ or 2-norm $||\bf{h}||_2$. Meanwhile, a threshold $\lambda_{RAUD}$ should be set to distinguish the $\bf{F}$ of UEs in different status. Different from the threshold $\lambda_{AUD}$ of AUD module, $\lambda_{RAUD}$ is related to SNR. As SNR increases, channel estimation error is reduced, and the $\bf{F}$ of inactive UEs will be much more tends to zero. Then, the value of $\lambda_{RAUD}$ should be reduced to adapt the new characteristic value. 
We do a lot of UL grant-free transmission tests for UEs with known status under different SNR, and obtain a collection of $\bf{F}$ characteristic values. 
By analyzing the size of $\bf{F}$ and their corresponding UE status, an empirical curve $\lambda_{RAUD}(SNR)$ related to SNR is obtained.

Then the following RAUD algorithm achieving the capability to identify UEs status can be developed according to the theory and analysis above. It can be seen that the algorithm has only a few norm and comparison operations, and the added complexity of the RAUD algorithm is very low compared to the entire UL grant-free SCMA receiver.

\begin{algorithm}[!h]
	\caption{RAUD algorithm} 
	\hspace*{0.02in} {\bf Input:} \\
	\hspace*{0.19in} Equivalent fading channel gain vector of potential pilots: \\
	\hspace*{0.19in}  $\{{{\bf{h}}_1}, {{\bf{h}}_2}, ... , {{\bf{h}}_P} \}$ \\
	\hspace*{0.19in} Potential UE list:  $\{1,...,P\}$ \\
	\hspace*{0.19in} Signal noise ratio: $SNR$\\
	\hspace*{0.02in} {\bf Output:} \\
	\hspace*{0.19in} Equivalent fading channel gain vector of active pilots:\\ \hspace*{0.19in} $\{{{\bf{h}}_1}, {{\bf{h}}_2}, ... , {{\bf{h}}_R} \}$ \\
	\hspace*{0.19in} Active UE list: $\{1,...,R\}$
	\begin{algorithmic}[1]
		\State $p = 1$ 
		\While{$p \le P$} 
		\State ${\bf{F}}_p = ||{\bf{h}}_p||_1 \ or \ ||{\bf{h}}_p||_2$
		\If{${\bf{F}}_p \geq {{\bf{\lambda}}_{RAUD}(SNR)}$} 
		\State $p \to$ Active UE list
		\Else
		\State Delete ${{\bf{h}}_p}$ in $\{{{\bf{h}}_1}, {{\bf{h}}_2}, ... , {{\bf{h}}_P} \}$
		\EndIf 
		\State $p = p + 1$
		\EndWhile
	\end{algorithmic}
\end{algorithm}

The improved UL grant-free SCMA receiver is depicted in Fig. 3. For the RAUD module is a further operation on the potential UE list after AUD module, the improved receiver is also called two-step AUD receiver.
Fig. 2, which replaces the JMPA module with the MPA module, is called one-step AUD receiver. The operation flow of the two-step AUD receiver structure can be described in the following. AUD module use the received pilot signal to identity inactive UEs/pilots and reduce the number of potential active UEs/pilots from $K$ to $P$. Then CE module performs channel estimation to get $P$ equivalent fading channel plural vectors. Each potential active UE corresponds to one ${\bf{F}}$ characteristic value. $\bf{F}$ and $\lambda_{RAUD}(SNR)$ can be used to distinguish the status of UEs/pilots in RAUD module. Then, the length of potential UE list can be reduced from $P$ to $R$. Finally, MPA decoder decode the data of $R$ real active UEs.
\begin{figure}[!h]
	\centering
	\includegraphics[width=0.48\textwidth]{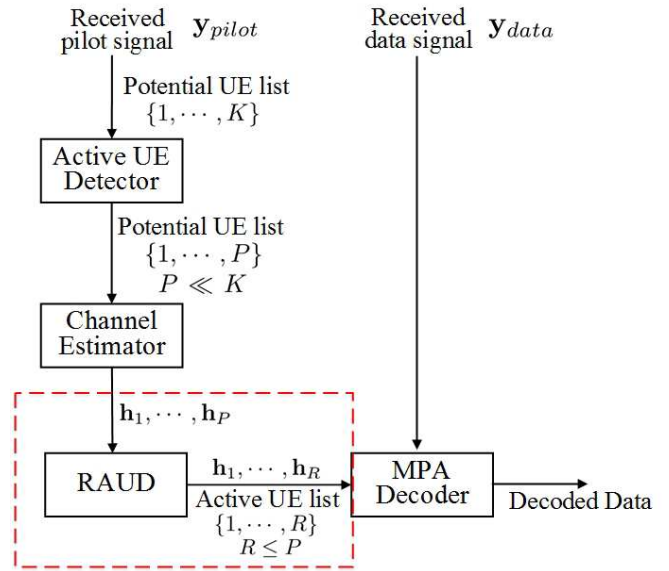}
	\vspace*{-10pt}
	\caption{UL grant-free SCMA two-step AUD receiver structure including active UE detector, channel estimator, RAUD module, and MPA decoder.}
	\label{3}
	\vspace*{-5pt}
\end{figure}

\section{SIMULATION RESULTS}

In the UL grant-free scenario, the miss detection probability and false alarm probability are important parameters for measuring receiver performance. The miss detection probability is defined as the ratio of the number of active UEs misinterpreted as inactive UEs to the total number of active UEs. Miss detection means that the loss of active UEs data. The false alarm probability is defined as the ratio of the number of inactive UEs regarded as active UEs to the total number of inactive UEs. High false alarm probability will lead to MPA decoding performance degradation and increased computational complexity.
In order to make the performance of two-step AUD receiver better than the one-step AUD receiver, we divide the function of the modules. The AUD module is mainly used to reduce the miss detection probability, while the RAUD module is aimed to reduce the false alarm probability without increasing miss detection probability.

Let us consider an UL grant-free SCMA system. The simulation parameters are shown in the Table I. Refer to Fig. 1, there are six different codebooks assigned to different groups in our simulation. The codebook in each group corresponds to three pilots. The length of pilot sequence is six resource blocks (RBs).

\begin{table}[!h]
	\centering
	\caption{Simulation Parameters}
	\footnotesize
	\begin{tabular}{ll}
		\toprule
		\textbf{Description}      & \textbf{Values} \\
		\hline
		\midrule
		Potential UEs                     & 18       \\ 
		Active  UEs                      & 6       \\ 
		The Number of Pilot Sequences    & 18 \\
		The Number of Codebooks			& 6  \\
		The length of Pilot Sequence		& 6RB  \\
		Channel model				& EPA/EVA\cite{b15}  \\
		AUD algorithm    			& FOCUSS  \\
		\bottomrule
	\end{tabular}
	\label{l2ea4-t1}
\end{table}

\subsection{Effect of false alarm probability}

High false alarm probability means that there are many inactive UEs enter the MPA decoder. On the one hand, these inactive UEs will interfere with the decoding of active user data. Fig. 4 shows the impact of different false alarm probability on BER performance under the EPA channel. Note that the BER performance deteriorates with the increase of the false alarm probability. One the other hand, inactive UEs could increase the computational complexity of MPA decoder, thereby increasing the transmission delay. The complexity of MPA decoding algorithm can be simply expressed by the Eq.(4) \cite{b16}.

\begin{equation}\label{4}
O(  {N_{iter} \sum\limits_{i = 1}^{SF} {  M_p^{d(i)} }})
\end{equation}

\noindent
where $N_{iter}$ is the number of iterations. $SF$ indicates the total number of the time-frequency resources of OFDMA used by the UEs. $M_p$ is the order of modulation. $d(i)$ indicates the number of UEs occupying the $i$-th resource. The increase of the number of inactive UEs leads to increased $d(i)$, which makes the computational complexity of MPA decoding algorithm increase exponentially. Therefore, reducing false alarm probability is necessary for the UL grant-free SCMA receiver.

\begin{figure}[!h]
	\centering
	\includegraphics[width=0.48\textwidth]{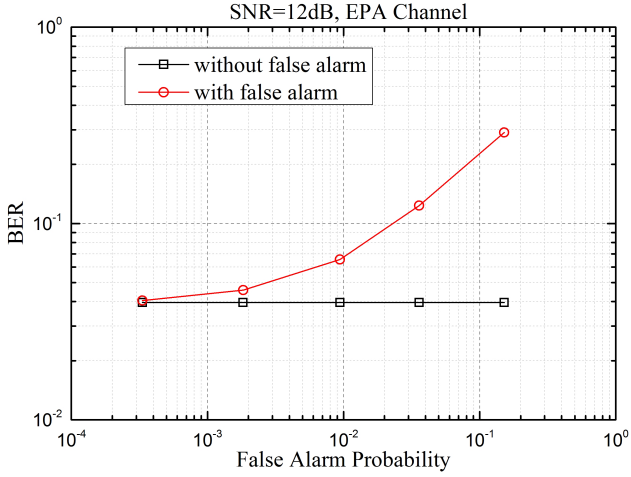}
	\vspace*{-10pt}
	\caption{Effect of false alarm probability on BER performance.}
	\label{4}
	\vspace*{-5pt}
\end{figure}

\subsection{The characteristic value of UEs}

Fig. 5 shows the probability distribution about normalized characteristic values of UEs.
The red and blue histogram represent inactive and active UEs respectively. Whether the AUD module or the RAUD module, their working principle is to use the difference between the characteristic values of active UEs and inactive UEs to identify the UEs' status. If there is no obvious difference between the characteristic values of the two status of UEs, it is difficult to set a threshold to accurately distinguish them. Comparing Fig. 5 (a) and Fig.5 (b), Fig. 5 (c) and Fig. 5 (d), the overlapping area of the one-step AUD receiver is larger than that of two-step AUD receiver under the EPA channel and EVA channel. Therefore, two-step AUD receiver can identify UEs' status more accuracy than one-step AUD receiver.
At the same time, we can see that when the receiver's channel changes from EPA to EVA, the overlapping area increases. That means both receivers will experience a drop in active UEs detection performance with the enhancement of channel frequency selection.

\begin{figure}[!h]
	\centering
	\includegraphics[width=0.38\textwidth]{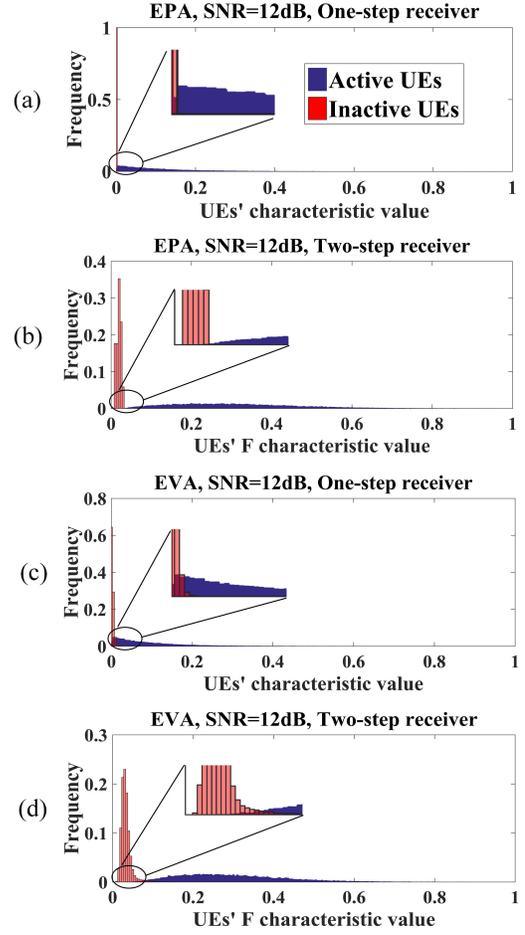}
	\vspace*{-5pt}
	\caption{Histogram of the UEs' characteristic value with/without RAUD module under EPA/EVA channel.}
	\label{5}
	\vspace*{-5pt}
\end{figure}
\begin{figure}[!h]
	\centering
	\includegraphics[width=0.50\textwidth]{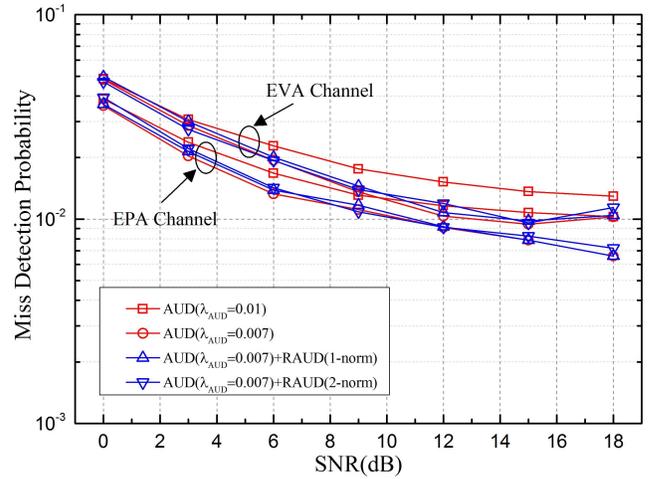}
	\vspace*{-20pt}
	\caption{Miss Detection Probability.}
	\label{6}
	\vspace*{-15pt}
\end{figure}
\begin{figure}[!h]
	\centering
	\includegraphics[width=0.50\textwidth]{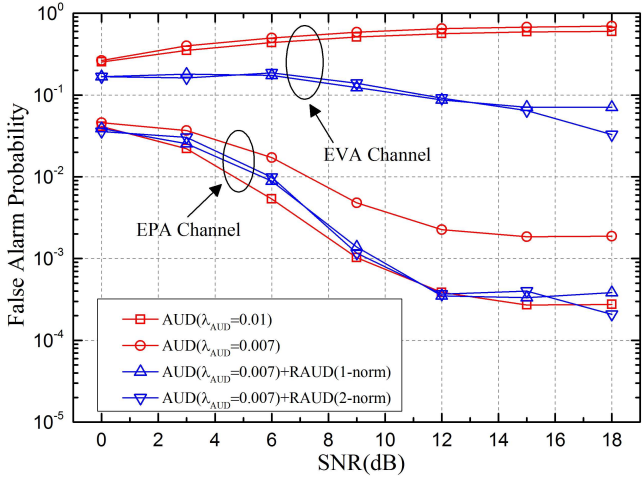}
	\vspace*{-10pt}
	\caption{False Alarm Probability.}
	\label{7}
	\vspace*{-6pt}
\end{figure}

\subsection{Performance comparison}

In order to make the two-step AUD receiver has a lower miss detection probability than the one-step AUD receiver. We reduce the threshold $\lambda_{AUD}$ of AUD module in the two-step AUD receiver from 0.01 to 0.007. As shown in Fig. 6, either under EPA channel or EVA channel, a lower value of $\lambda_{AUD}$ makes lower miss detection probability. But at the same time,  $\lambda_{AUD}$=0.007 makes a higher false alarm probability, as shown in Fig. 7. Aiming at this problem, the addition of the RAUD module can effectively reduce the false alarm probability, while almost not increasing the miss detection probability. Observing curve AUD($\lambda_{AUD}$=0.01) and curve AUD($\lambda_{AUD}$=0.007)+RAUD(1-norm) in Fig. 7. Compared to EPA channel, RAUD module reduces false alarm probability more obviously under EVA channel. It also can be seen that the RAUD algorithm has the same performance whether 1-norm or 2-norm is used.

From the figures above, it can be confirmed that the two-step AUD receiver can bring lower miss detection probability and false alarm probability than one-step AUD receiver. Meanwhile, the RAUD module can reduce the computational complexity of MPA decoder and optimize the decoding performance. Actually, the RAUD module can more effectively reduce the false alarm probability under EVA channel and improves the adaptability of UL grant-free SCMA receiver in frequency selective channel.

\section{CONCLUSION}
In this paper, we introduce the transmitter and the original receiver of SCMA multiple access in the UL grant-free scenario, and analyze the principle, advantages and disadvantages of AUD, CE and JMPA module. 
Then, we propose a two-step AUD receiver scheme contains RAUD module. By making full use of the channel gain and analyzing the difference between characteristic values of the two status of UEs, the RAUD module can selected out the inactive UEs that AUD module misinterpreted.
Finally, we verify that the lower false alarm probability is beneficial to improve the decoding performance and reduce the computational complexity of MPA decoder. The simulation results show that the two-step AUD receiver has lower miss detection probability and false alarm probability than the one-step AUD receiver, especially under the frequency selective channel.
From analysis and simulation results, it is confirmed that the proposed two-step AUD receiver provide a way of designing an UL grant-free SCMA system for adapting the frequency selective channel.

\section*{Acknowledgment}
This work is supported by the China Natural Science Funding (NSF) under Grant 61671089, Huawei Cooperation Project.


\end{document}